\documentclass[conference]{IEEEtran}
\IEEEoverridecommandlockouts

\usepackage{cite}
\usepackage{amsmath,amssymb,amsfonts}
\usepackage{graphicx}
\usepackage{textcomp}
\usepackage{xcolor}
\def\BibTeX{{\rm B\kern-.05em{\sc i\kern-.025em b}\kern-.08em
    T\kern-.1667em\lower.7ex\hbox{E}\kern-.125emX}}

\usepackage{cleveref}
\usepackage{tabularray}
\usepackage{booktabs}
\usepackage{multirow}
\usepackage{makecell}

\usepackage{tabularray}
\usepackage{booktabs}
\usepackage{multirow}
\usepackage{makecell}

\usepackage{algorithm}
\usepackage[noend]{algpseudocode}
\makeatletter
\def\BState{\State\hskip-\ALG@thistlm}
\makeatother
\newcommand{\E}{\mathbb{E}}

\newcommand{\R}{\mathbb{R}}

\newtheorem{definition}{Definition}

\usepackage{matlab-prettifier}
\usepackage{color}
\usepackage{mathrsfs}
\usepackage[shortlabels]{enumitem}
\usepackage{subcaption}

\begin{document}



\title{\LARGE \bf When to Deceive: A Cross-Layer  Stackelberg Game Framework for Strategic Timing of Cyber Deception
}

\author{
Ya-Ting Yang  and  Quanyan Zhu\\
Department of Electrical and Computer Engineering, 
Tandon School of Engineering, \\ New York University, 
Brooklyn, NY, USA; 
\texttt{\{yy4348, qz494\}@nyu.edu}
\thanks{This work has been submitted to the IEEE for possible publication.
Copyright may be transferred without notice, after which this version may no longer be accessible.}
}

\maketitle

\begin{abstract}
Cyber deception is an emerging proactive defense strategy to counter increasingly sophisticated attacks such as Advanced Persistent Threats (APTs) by misleading and distracting attackers from critical assets. However, since deception techniques incur costs and may lose effectiveness over time, defenders must strategically time and select them to adapt to the dynamic system and the attacker’s responses. In this study, we propose a Stackelberg game-based framework to design strategic timing for cyber deception: the lower tactical layer (follower) captures the evolving attacker-defender dynamics under a given deception through a one-sided information Markov game, while the upper strategic layer (leader) employs a stopping-time decision process to optimize the timing and selection of deception techniques. We also introduce a computational algorithm that integrates dynamic programming and belief-state updates to account for the attacker’s adaptive behavior and limited deception resources. Numerical experiments validate the framework, showing that strategically timed deceptions can enhance the defender’s expected utility and reduce the risk of asset compromise compared to baseline strategies.
\end{abstract}


\section{Introduction}
Advanced Persistent Threats (APTs) \cite{8606252} have become a significant challenge for enterprise networks due to their stealth, persistence, and sophisticated attack methods. As detailed in the MITRE ATT\&CK framework \cite{strom2018mitre}, the tactics and techniques of APTs are becoming increasingly complex, necessitating the development of more advanced and proactive defense strategies to complement existing security measures. 

\begin{figure}
    \centering
    \includegraphics[width=0.95\linewidth]{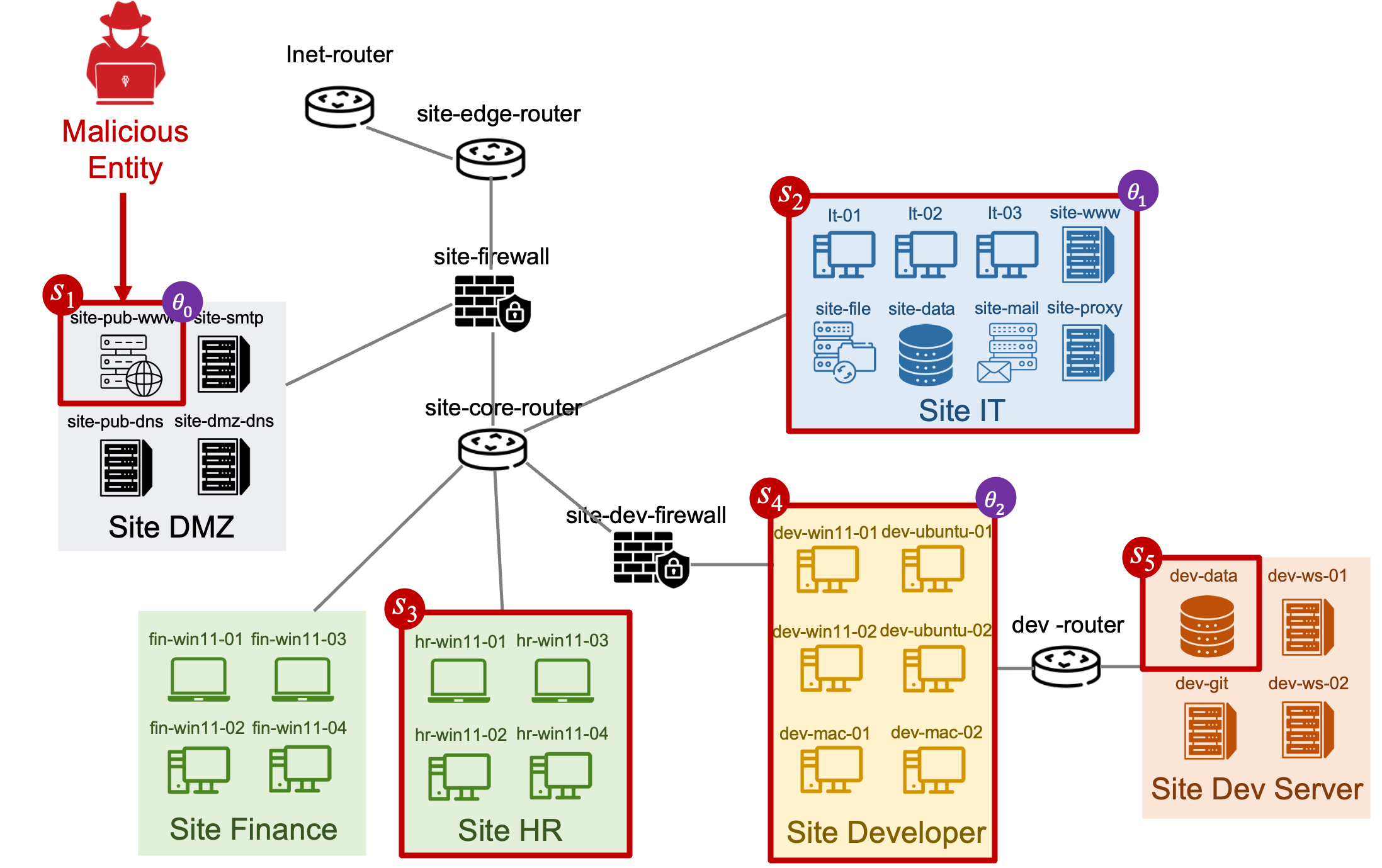}
    \caption{An example enterprise network and attack path. The path contains 5 steps (web server, site IT, site HR, site developer, and critical asset). The path starts from the web server, which is open to the external network, and the defender seeks to optimize the timing and selection of deception.}
    \label{fig:net}
\vspace{-5mm}
\end{figure}

Cyber deception is an emerging proactive defense strategy to counter increasingly sophisticated attacks. Since attackers often operate with incomplete information about the system, defenders can exploit this vulnerability by manipulating the attacker’s perception through the strategic deployment of deception technologies \cite{li2024symbiotic}. Deceptions such as Moving Target Defense (MTD) \cite{jajodia2011moving}, decoy files, fake data paths, and honeypots \cite{javadpour2024comprehensive} introduce uncertainty and misinformation, misleading the attacker to non-critical resources and diverting their attention away from critical assets. 
For example, in Fig. \ref{fig:net}, the defender can use deceptions $\theta_0, \theta_1, \theta_2$ at various points along the attack path, $s_1, s_2, s_4$, to delay or misdirect the attacker, protecting assets located on the developer’s server.

However, an attacker may be deceived by a particular technique once, but once the deception is recognized, the attacker will not be fooled again unless the defender employs a different deception method. As a result, previously deployed or activated deception techniques cannot be reused. Moreover, implementing deception techniques can incur significant costs, making their unrestricted use impractical. Therefore, the defender must strategically choose the right deception technique at the right time, making the deception design challenge into a timing problem. Another challenge that remains is the dynamic nature of the system, where both the system environment and the attacker’s behavior continuously evolve. This requires deception strategies to be adaptable, evolving in response to changes within the system as well as the attacker’s perceptions and responses to maintain the defender's strategic advantages.

The conflict of interest between the defender and the attacker naturally leads to a game-theoretic approach for deception design \cite{nan2019behavioral,anwar2022honeypot}. Hence, to address the aforementioned challenges, we propose a Stackelberg game \cite{cranford2018learning}-based framework in this study, designed to analyze how the defender can strategically time the deployment of deception techniques to maintain a strategic advantage throughout the attacker's lifetime $K$ while protecting critical assets. The ``lower'' tactical layer is modeled as a one-sided information Markov game \cite{bacsar1998dynamic}, capturing the interactions between the defender and the attacker within the dynamic system for a given deception technique and its configuration. In this setup, the defender has full access to system information, while the attacker possesses only limited knowledge. Motivated by the observation that different deception techniques can delay an attacker at specific stages along the attack path for varying durations \cite{ferguson2018tularosa}, we formulate the “upper” strategic layer as a stopping-time decision process. This layer captures the strategic planning of deception deployment, where deception techniques are treated as limited resources. In this context, both the timing and selection of deception techniques are critical to sustaining the defender’s strategic advantage. Effective upper-layer planning requires anticipating not only how the defender will utilize each technique but also how the attacker is likely to respond to them at different points in time. Importantly, the success of deceiving an adaptive attacker under resource constraints depends heavily on the foresight and coordination embedded in this higher-level planning process.

Moreover, the interdependence between the upper and lower layers is fundamental: decisions at the upper strategic level (i.e., timing $k$ to activate which deployed deception $\theta$) influence, and are influenced by, the tactical interactions modeled in the lower layer. This coupling motivates the use of a bi-level Stackelberg game formulation. The upper layer acts as the leader, guiding the deployment of deception over time, while the lower layer models the real-time interactions between the defender and the attacker under asymmetric information.

To this end, our contribution can be summarized as follows. First, we develop a Stackelberg game-theoretic framework for the co-design of strategic planning and tactical implementation of cyber deception. The lower, tactical layer captures the evolving attacker-defender dynamics through a one-sided information Markov game, enabling the analysis of behavioral responses under different deception configurations. In turn, the upper, strategic layer formulates a stopping-time decision process that guides the optimal timing and selection of deception techniques, subject to resource constraints and the attacker’s evolving belief dynamics.
%
Second, we introduce a computational algorithm that integrates dynamic programming with belief-state updates to compute the equilibrium strategies in the tactical layer and to determine the optimal switching strategy in the strategic layer. 
Third,  we validate the proposed framework through numerical case studies conducted in an enterprise network, which reveal that strategic deception timing can improve the defender’s expected utility and reduce the likelihood of critical asset compromise by $25\sim 50\%$ compared to baseline strategies without switching.

\section{Problem Formulation}

We consider an enterprise network consisting of routers, different sites, servers, users, and critical assets, as illustrated in Fig. \ref{fig:net}.  
Typically, attackers who launch APT initiate their intrusion from the external network, often targeting the web server as the first point of entry. An example of the attack path that an attacker might exploit to access the company’s critical assets is illustrated in Fig. \ref{fig:net}.

The proposed Stackelberg game-based framework consists of two layers. The lower tactical layer captures the interactions between the attacker and the defender with the dynamic system environment, while the upper strategic layer assists the defender in deciding when to activate which deployed deception techniques along with its associated configuration. 

\subsection{The Lower Tactical Layer}
The tactical layer focuses on the interaction between the defender (she) and the attacker (he) given a specific deception technique. Since the attacker neither knows whether there is deception nor which deception is in use, we model this interactive dynamic decision-making between the defender and the attacker using a Markov game with one-sided information, where the defender possesses complete access to the game information, while the attacker remains unaware of the game mode representing the activated deception technique. The game unfolds in discrete time, progressing through stages denoted by $k = 0, 1, \cdots, K$, with $K$ being the last stage of the attacker's lifetime within the system.

\begin{definition}[Tactical Layer Problem (TLP)]
\label{def:BG}
    The Tactical Layer Problem (TLP) is defined as a set of Markov Games $\Phi_\Theta=\{\Gamma^{\theta}\}_{\theta \in \Theta}$ with one-sided information between the defender and the attacker, where $\Theta$ represents the set of possible game modes. The game $\Gamma^\theta$ can be represented by a tuple $\Gamma^\theta=\langle \mathcal{N}, \mathcal{S}, \theta, \{\mathcal{A}_i\}_{i \in \mathcal{N}}, T^\theta, \{r_i^\theta\}_{i \in \mathcal{N}}, \gamma, \{\mathcal{I}_i\}_{i \in \mathcal{N}} \rangle$, where each component represents:
    \begin{itemize}
        \item $\mathcal{N}=\{d, a\}$ represents the set of players, where $P_d$ represents the defender and $P_a$ denotes the attacker. 
        \item $\mathcal{S}$ is the finite set of all possible states (such as location within the system), and the true state $s \in \mathcal{S}$ is directly observable for both $P_d$ and $P_a$.
        \item $\theta \in \Theta$ is the game mode (such as which deception technique is deployed and activated), which is known to $P_d$ but unknown to $P_a$.
        \item $\mathcal{A}_i$ represents the action set for player $P_i, i \in \mathcal{N}$.
        \item $T^\theta: \mathcal{S} \times \mathcal{A}_d \times \mathcal{A}_a \mapsto \Delta(\mathcal{S})$ is the state transition function conditional on current state and actions under game mode $\theta$, with $T^\theta(s'|s, a_d, a_a)$ representing the probability of transitioning to the next state $s' \in \mathcal{S}$ given the current state $s \in \mathcal{S}$, the defender's action $a_d \in \mathcal{A}_d$, and the attacker's action $a_a \in \mathcal{A}_a$.
        \item $r^\theta_i:\mathcal{S} \times \mathcal{A}_d \times \mathcal{A}_a \mapsto \mathbb{R}$ is the immediate reward function for player $P_i$ when actions $a_d \in \mathcal{A}_d$ and $a_a \in \mathcal{A}_a$ are taken in state $s \in \mathcal{S}$ with game mode $\theta \in \Theta$.
        \item $\gamma \in [0, 1]$ is the discount factor that determines the importance of future rewards.
        \item $\mathcal{I}_i$ is the information structure, where $\mathcal{I}_i^k$ represents the information available to Player $P_i$ at stage $k$.
    \end{itemize}
\end{definition}
The game starts in an initial state $s^0 \in \mathcal{S}$. At stage $k$ with state $s^k$, the players $P_d$ and $P_a$ select actions $a_d^k \in \mathcal{A}_d$ and $a_a^k \in \mathcal{A}_a$ based on their information structure $\mathcal{I}_d^k$ and $\mathcal{I}_a^k$, respectively. Then, the next state $s^{k+1} \sim T^\theta(s^k, a_d^k, a_a^k)$ is determined by the transition function associated with the current game mode $\theta$. At the same time, player $P_d$ receives $r_d^\theta(s^k, a_d^k, a_a^k)$ while player $P_a$ receives $r_a^\theta(s^k, a_d^k, a_a^k)$.

To optimize their strategy against the opponent, players can leverage all available information up to the point of decision-making, which is known as the behavioral strategy. Hence, at stage $k$, each player's strategy is a mapping from the information structure to the distribution of actions: $\pi_i^k:\mathcal{I}_i^k \mapsto \Delta(\mathcal{A}_i), i \in \mathcal{N}$, with $\pi_i = [\pi_i^0, \cdots, \pi_i^k, \cdots, \pi_i^{K}], i \in \mathcal{N}$. In this work, we specifically focus on the Markov (mixed) strategy, a particular type of behavioral strategy, for both players. Since the defender $P_d$ has complete information, in the Markov setting, $\mathcal{I}_d^k=\{s^k, \theta\}$, then $\pi_d^k: \mathcal{S} \times \Theta \mapsto \Delta(\mathcal{A}_d)$. However, due to unawareness of the deception technique, only the true state $s^k$ is directly observable to the attacker, $\mathcal{I}_a^k=\{s^k\}$, then $\pi_a^k: \mathcal{S} \mapsto \Delta(\mathcal{A}_a)$. Note that the true game mode $\theta$ is not directly observable to the attacker, the attacker in this case can only maintain a belief of the current game mode $b^k \in \Delta(\Theta)$, with $b^k(\theta)$ representing the attacker's belief of being in mode $\theta \in \Theta$ at stage $k$. A typical method for an attacker to update his belief is through Bayes' rule, as in \eqref{eq:b_update}.
\begin{figure*}
\begin{equation}
        b^{k+1}(\theta|s^{k+1}) = \frac{\sum_{a_d, a_a} T^{\theta}(s^{k+1}|s^k, a_d, a_a) \pi_d^k(a_d|s^k, \theta)\pi_a^k(a_a|s^k)b^k(\theta|s^k)}{\sum_{a_d, a_a, \theta'} T^{\theta'}(s^{k+1}|s^k, a_d, a_a) \pi_d^k(a_d|s^k, \theta')\pi_a^k(a_a|s^k)b^k(\theta'|s^k)}.
    \label{eq:b_update}
\end{equation}\\
\vspace{-5mm}
\end{figure*}
Then, given the initial state $s^0 \in \mathcal{S}$ and both players' TP $\pi=[\pi_d, \pi_a$], player $P_d$ aims to maximize her expected cumulative discounted reward over a finite horizon $K$:
\begin{equation}
    R_d(s^0, \pi_d, \pi_a, \theta) = \E_{\pi_d, \pi_a, T^\theta} \left[\sum_{k=0}^K \gamma^k r_d^\theta(s^k, a_d^k, a_a^k)\right].
\label{eq:e_reward_d}
\end{equation} 
As the attacker does not know the true game mode $\theta$, he needs to take the expectation of his cumulative discounted reward over a finite horizon $K$ with respect to $\theta$ at each stage $k$. That is, player $P_a$ intends to maximize:
\begin{equation}
    R_a(s^0, \pi_d, \pi_a, b^0) = \E_{\pi_d, \pi_a, T^\theta, b} \left[\sum_{k=0}^K \gamma^k r_a^\theta(s^k, a_d^k, a_a^k)\right],
\label{eq:e_reward_a}
\end{equation} where $b^0$ is the attacker's prior belief and $b=[b^0, \cdots, b^K]$ is a sequence of beliefs updated according to \eqref{eq:b_update}.

Since the state transition functions, which define the probability of transitioning to the next state given the current state and actions, vary across different game modes, the defender acts not only as a player but also as a system configurator. In this context, the defender is then responsible for deploying deception methods by selecting which $\theta \in \Theta$ to activate. 

\subsection{The Upper Strategic Layer}
Given the tactical decision-making at the lower layer, which analyzes the effectiveness of each deception technique, the upper strategic layer focuses on optimizing the timing and selection of these techniques for activation by the defender. It is worth noting that once an attacker recognizes a deception, the technique cannot be reused. Additionally, deception methods such as honeypots incur costs, making unrestricted use impractical. Given these limitations, we assume that the defender has a limited budget for activating deception techniques to prevent the attacker from reaching critical assets. In this context, the defender can take actions from a set of deception techniques, $\Theta=\{\theta_0, \theta_1, \cdots, \theta_{|\Theta|-1}\}$, leading to switching for the system configuration (i.e., resulting in changes of the transition probability) up to a budget of $M$ times within $K$ stages, with $M \leq K$ and each action can only be taken once. For example, in Fig. \ref{fig:net}, suppose $\theta_0$ represents the absence of deception and $M=2$, the upper strategic layer decides whether and when to activate deception at site IT and which of $\theta_1, \theta_2$ to choose. If $\theta_1$ is selected at this stage, the upper strategic layer can only decide when to activate $\theta_2$ at the other sites later on.

\begin{definition}[Strategic Layer Problem (SLP)]
\label{def:CAP}
    Given the TLP $\Phi_\Theta$ defined in Definition \ref{def:BG}, the Strategic Layer Problem (SLP) for the defender can be represented by a tuple $\Sigma(\Phi_\Theta)=\langle \Theta,  K, M, \mathcal{X}, \{U^\theta\}_{\theta \in \Theta}\rangle$, where each element represents:
    \begin{itemize}
        \item $\Theta=\{\theta_0, \theta_1, \cdots, \theta_{|\Theta|-1}\}$ is the action set for possible deception techniques of the defender. The set of actions already taken is represented by $\Theta' \subseteq \Theta$.
        \item $K$ is the total stages of the problem, representing the attacker's lifetime within the system.
        \item $M$ is the budgeted number of times to switch (i.e., activate a deployed deception technique). The number of switches remaining is denoted as $m$.
        \item $\mathcal{X}$ is the state space, each state $x^k=(s^k, b^k)$ at stage $k$ consists of the state of TLP $s^k$ and the attacker's belief $b^k$ updated according to \eqref{eq:b_update}.
        \item $\eta$ is a threshold for the attacker's belief.
        \item $\{U^\theta\}_{\theta \in \Theta}: \mathcal{X} \times \mathcal{A}_d \times \mathcal{A}_a \mapsto \R$ are the utility functions for the defender, with 
        \begin{equation}
        \begin{aligned}
            U^\theta(x^k, a^k_d, a^k_a) &= U^\theta(s^k, b^k, a^k_d, a^k_a)\\ &= r_d^\theta(s^k, a^k_d, a^k_a)\boldsymbol{1}_{b^k(\theta) \leq \eta},
        \end{aligned}
        \label{eq:payoff}
        \end{equation} where $\eta$ is a threshold for the attacker's belief of the current game mode $\theta^k$ at stage $k$.
    \end{itemize}
\end{definition}
The defender's objective is to maximize her utility by deciding the right time to switch and the right action to take (the right deception technique to activate). Let $\bar{k}=[k^0, k^1, \cdots, k^m]$ represent the initial stage and stages at which switches occur, and let $\bar{\theta}$ denote the corresponding deception techniques to be switched to. Then, given any tactical profile (TP) $\pi=[\pi_d, \pi_a]$ the defender's problem is:
\begin{equation}
    \begin{aligned}
        \max_{\bar{k},\bar{\theta}} \sum_{m=0}^{M-1} \E_{\pi_d, \pi_a, T^{\theta^m}}\bigg[\sum_{k=k^{m}}^{k^{m+1}}U^{\theta^m}(x^k, a_d^k, a_a^k)\bigg].
    \end{aligned}
\end{equation}

To this end, we have introduced  the components of the framework for deception design. The equilibrium at the tactical layer and the optimal switching strategy for the strategic layer will be detailed in the next section.

\section{Equilibrium and Optimal Timing}
\subsection{Equilibrium at the Tactical Layer}
At the tactical layer, the defender and attacker aim to maximize their expected cumulative utilities, \eqref{eq:e_reward_d} and \eqref{eq:e_reward_a}, within the finite horizon $K$ by designing their strategies, $\pi_d$ and $\pi_a$, respectively. Since the utilities depend on the actions of both players, the one-sided $\epsilon$-Perfect Bayesian Nash Equilibrium ($\epsilon$-PBNE) naturally arises as the solution concept.
\begin{definition}[One-sided $\epsilon$-PBNE]
    Consider the TLP $\Phi_\Theta$ defined in Definition \ref{def:BG}. Given $\epsilon \in \mathbb{R}_{\ge 0}$, a tactical profile (TP) $\pi^*=[\pi_d^*, \pi_a^*]$ constitutes a one-sided $\epsilon$-Perfect Bayesian Nash Equilibrium ($\epsilon$-PBNE) if it satisfies:\\
    (C1) Belief Consistency: under TP $\pi^*=[\pi_d^*, \pi_a^*]$, player $P_a$'s belief $b^k$ at each stage $k=0, \cdots, K$ satisfies \eqref{eq:b_update}.\\
    (C2) Sequential Rationality: For all given game mode $\theta$ and initial state $s^{k_0} \in \mathcal{S}$ at every initial stage $k_0 \in \{0, \cdots, K\}$, each TP $\pi_i^{*, k_0:K}=[\pi_i^{*,k_0}, \cdots, \pi_i^{*,K}], i \in \mathcal{N}$ must be $\epsilon$-optimal in expectation, 
    \begin{align}
        &R_d(s^{k_0}, \pi_d^{*, k_0:K}, \pi_a^{*, k_0:K}, \theta) + \epsilon \nonumber\\
        &\qquad \geq R_d(s^{k_0}, \pi_d^{k_0:K}, \pi_a^{*, k_0:K}, \theta), \forall \pi_d^{k_0:K} \in \Pi_d^{k_0:K},\\
        &R_a(s^{k_0}, \pi_d^{*, k_0:K}, \pi_a^{*, k_0:K}, b^{k_0}) + \epsilon \nonumber\\
        &\qquad \geq R_a(s^{k_0}, \pi_d^{*, k_0:K}, \pi_a^{k_0:K}, b^{k_0}), \forall \pi_a^{k_0:K} \in \Pi_a^{k_0:K},
    \end{align} where $\Pi_i^{k_0:K}$ denotes the set of all TPs for player $P_i$ from stage $k_0$ to $K$.
    \label{def:PBNE}
\end{definition} Note that when $\epsilon=0$, the TP $\pi^*=[\pi_d^*, \pi_a^*]$ is an exact PBNE. Then, following studies \cite{huang2020dynamic,ge2023gazeta}, the one-sided $\epsilon$-PBNE can be found by dynamic programming and backward induction.

\subsection{Optimal Timing at the Strategic Layer} \label{sec:optimal_switch}
At the strategic layer, we denote $\bar{x}^k=(x^k, \theta^k, m, \Theta')$ as the extended state space at stage $k$ with current game mode $\theta^k$, with $x^k=(s^k, b^k)$ and $m$ being the number of remaining switches (to activate another unused deception technique).  
Then, the optimal switching strategy can be solved using dynamic programming by considering the problem backward from the final stage $k=K$. Given any tactical profile (TP), $\pi=[\pi_d, \pi_a]$ and belief sequence $b$ updated according to \eqref{eq:b_update} for  the TLP, the value function for the defender at stage $K$ is simply the terminal utility $U_T:\mathcal{S} \mapsto \mathbb{R}$, stating whether the attacker reaches the state leading to severe loss or containing the critical asset; i.e.,
\begin{equation}
    \begin{aligned}
        &V_\pi^K(\bar{x}^K)=U_T(s^K).
    \end{aligned}
\label{eq:v_K}
\end{equation}

As for stage $k < K$, if there are no remaining times to switch ($m=0$), the defender must stay and remain with the current deception technique. However, if there are remaining times to switch ($m>0$), the defender can either continue with the current technique or switch to another unused one, depending on which option offers a higher value for her. Therefore, we have the value function at stage $k$ given the defender and the attacker's TP $\pi=[\pi_d, \pi_a]$ and belief sequence $b$ as follows:
\begin{equation}
        V_\pi^k(\bar{x}^k)=\begin{cases}
            V^k_{\pi,stay}(\bar{x}^k), &\text{ if } m=0,\\
            \max(V^k_{\pi,stay}(\bar{x}^k), V^k_{\pi,switch}(\bar{x}^k)), &\text{ if } m>0,
        \end{cases}
\label{eq:v_k}
\end{equation} with the value of staying being
\begin{equation*}
    \begin{aligned}
    &V^k_{\pi,stay}(\bar{x}^k) =V^k_{\pi,stay}(s^k, b^k, \theta^k, m, \Theta')\\& =\sum_{a_d \in \mathcal{A}_d}\sum_{a_a \in \mathcal{A}_a}\pi_d^k(a_d^k|s^k, \theta^k)\pi_a^k(a_a^k|s^k)\big[U^\theta(s^k, b^k, a_d^k, a_a^k)\\& \quad +\sum_{s' \in \mathcal{S}}T^{\theta^k}(s'|s^k, a_d^k, a_a^k)V_\pi^{k+1}(s', b^{k+1}, \theta^k, m, \Theta')\big],
\end{aligned}  
\label{eq:v_stay}
\end{equation*} where $T^{\theta^k}$ is the transition probability associated with the current mode $\theta^k$, and the attacker's belief $b^{k+1}$ is consistent with TP from the tactical layer and updated according to \eqref{eq:b_update}. The value at the extended state $\bar{x}^k$ describes the expected return starting from that state and then acting according to the given TP when the defender decides to stay in the current game mode. For the value of switching, 
\begin{equation*}
    \begin{aligned}
    &V^k_{\pi,switch}(\bar{x}^k) =V^k_{\pi,switch}(s^k, b^k, \theta^k, m, \Theta')=\\& \max_{\theta \in \Theta\setminus\Theta'} \sum_{a_d \in \mathcal{A}_d}\sum_{a_a \in \mathcal{A}_a}\pi_d^k(a_d^k|s^k, \theta)\pi_a^k(a_a^k|s^k)\big[U^\theta(s^k, b^k, a_d^k, a_a^k) \\& +\sum_{s' \in \mathcal{S}} T^\theta(s'|s^k, a_d^k, a_a^k)  V_\pi^{k+1}(s', b^{k+1}, \theta, m-1, \Theta'\cup\{\theta\})\big], 
\end{aligned}  
\label{eq:v_switch}
\end{equation*} where the $\max$ operator is used to identify the unused deception technique that can maximize the expected return starting from the current extended state, considering the subsequent actions that follow the given TP. 
The execution of the proposed framework is summarized in Algorithm \ref{algo:exe}.

\begin{algorithm}
  \caption{Deception Timing Design Execution} \label{algo:exe}
  \begin{algorithmic}[1]
    \State\textbf{Input} $\Phi_\Theta,\Sigma(\Phi_\Theta)$, TP $\pi$, consistent belief $b$, \\ $\ \ \qquad$ threshold $\epsilon, \eta \geq 0$
    \State \textbf{At} stage $k$, given $\bar{x}^k=(s^k$, $b^k$, $\theta^k$, $m$, $\Theta')$
    \State \textbf{Compute} $V_{\pi}^k(\bar{x}^k)$ using \eqref{eq:v_k}, decide $\theta^{k+1}$
    \If{$\theta^{k+1} \neq \theta^k$}
        \State  $m=m-1, \Theta'\cup\{\theta^{k+1}\}$
    \EndIf
    \State \textbf{Take} one step forward: $a_d^k \sim \pi^{k}_d, a_a^k \sim \pi^{k}_a$ 
    \State \textbf{Observe} next state: $s^{k+1} \sim T^{\theta^{k+1}}$ 
    \State \textbf{Update} belief $b^{k+1}(\theta|s^{k+1}k, a_d^k, a_a^k)$ according to \eqref{eq:b_update_exe}
    \State \Return $\bar{x}^{k+1} = (s^{k+1}, b^{k+1}, \theta^{k+1}, m, \Theta')$
  \end{algorithmic}
\end{algorithm}

During execution, the belief is updated ex-post, meaning it is computed after the action is executed and the state transitions, as shown in \eqref{eq:b_update_exe}. 
\begin{equation}
    \begin{aligned}
        &b^{k+1}(\theta|s^{k+1}, a_d^k, a_a^k)\\& = \frac{T^{\theta}(s^{k+1}|s^k, a_d, a_a) \pi_d^k(a_d|s^k, \theta)\pi_a^k(a_a|s^k)b^k(\theta|s^k)}{\sum_{\theta'} T^{\theta'}(s^{k+1}|s^k, a_d, a_a) \pi_d^k(a_d|s^k, \theta')\pi_a^k(a_a|s^k)b^k(\theta'|s^k)}.
    \end{aligned}
    \label{eq:b_update_exe}
\end{equation}
At stage $k$, given the current state $s^k$, belief $b^k$, mode $\theta^k$, remaining budgeted switches $m$, and the set of used techniques $\Theta'$, the strategic layer decides whether to switch modes ($\theta^{k+1}\neq \theta^k$) or maintain the current mode ($\theta^{k+1}=\theta^k$) based on the given TP $\pi$ obtained from the tactical layer. Then, a step is then executed according to the given TP, where actions $a_d^k \sim \pi^{k}_d, a_a^k \sim \pi^{k}_a$  are chosen, and the system transitions to a new state $s^{k+1}\sim T^{\theta^{k+1}}(s^k, a_d^k, a_a^k)$ according to the mode $\theta^{k+1}$ decided at the strategic layer.

\section{Numerical Results}


\subsection{Setups}
Consider the scenario where a software development company or enterprise seeks to enhance the security of its network systems and protect critical assets through deception design. The network topology of the enterprise and an example attack path that a red-team agent may exploit is shown in Fig. \ref{fig:net}, with the critical assets located on the developers' server site.  At the tactical layer, the red-team agent may select actions such as exploitation, obfuscation, and privilege escalation, while the defender may implement defenses like detection, access control, and monitoring. Each of these actions involves a sequence of specific techniques and procedures. At the strategic layer, the defender may choose the timing and the deception technique to activate, including banners with misinformation, decoy files, honeypots, etc. In this case, each element of the TLP $\Phi_\Theta$ and the SLP $\Sigma(\Phi_\Theta)$ are specified as follows.

\begin{figure}
    \centering
    \includegraphics[width=2.5in]{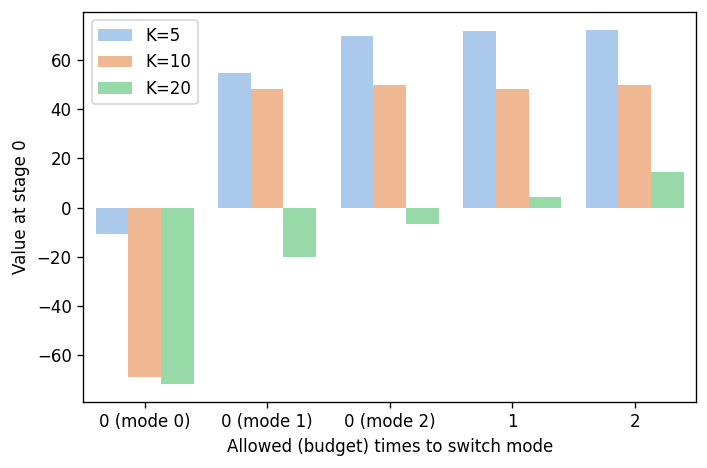}
    \caption{The results show the value at the initial stage for different budgeted switch times and varying attacker lifetimes. Three cases are analyzed for different initial game modes (modes 0, 1, and 2) when the budget is zero, while optimal switching is considered when the budget is non-zero.}
    \label{fig:v0}
\vspace{-5mm}
\end{figure}

\begin{itemize}
    \item $s \in \mathcal{S} = \{s_1, s_2, s_3, s_4, s_5\}$, where the states correspond to the attack steps in the attack path illustrated in Fig. \ref{fig:net}.
    \item $\theta \in \Theta = \{0, 1, 2\}$, where $\theta = 0$ represents the absence of deception, $\theta = 1$ denotes a basic deception technique, such as a misinformed banner, and $\theta = 2$ represents a more sophisticated deception, such as decoy files.
    \item $a_d \in \mathcal{A}_d = \{0, 1, 2\}$, 
    where $a_d = 0$, $a_d = 1$, and $a_d = 2$ correspond to the defense efforts (for tactics), such as strong access controls and behavioral analytics, required for $\theta = 0$, $\theta = 1$, and $\theta = 2$, respectively.
    \item $a_a \in \mathcal{A}_a = \{0, 1, 2\}$, where $a_a = 0$, $a_a = 1$, and $a_a = 2$ correspond to the attack efforts (for tactics), such as standard exploitation and privilege escalation, required for $\theta = 0$, $\theta = 1$, and $\theta = 2$, respectively.
\end{itemize} The detailed setups for the immediate rewards $r^\theta_d, r^\theta_a$, and the transition function $T^\theta$ for different game modes $\theta \in \Theta$ are summarized in Table \ref{tab:IRsetup}, and \ref{tab:TPsetup}, respectively.

For the strategic layer $\Sigma(\Phi_\Theta)$, the set of deception techniques $\Theta$, the state space $\mathcal{X}$, where $x^k =(s^k, b^k) \in \mathcal{X}$ follows from the TLP $\Phi_\Theta$. The total stage is set to $K=5, 10, 20$, and the budgeted switch is set to $M=0, 1, 2$. The payoffs $\{U^\theta\}_{\theta \in \Theta}$ follows \eqref{eq:payoff}, while the terminal rewards $U_T$ are summarized in Table \ref{tab:TRsetup}. The results for the value at the initial stage under optimal switching strategy, considering different budgeted times for switching and varying attacker lifetimes, are shown in Fig. \ref{fig:v0}.

\begin{table}[htbp]
\caption{Immediate Rewards}
\begin{center}
\begin{tabular}{ccc c}
\toprule
$a_d^k$ & $a_a^k$ & condition & $r_d^\theta(s^k, a_d^k, a_a^k)$ \\
\midrule
$\theta^k$ & \thead[l]{not $\theta^k$ \\ $\theta^k$}  & & \thead[c]{$10.0$ \\ $5.0$} \\
\midrule
not $\theta^k $ & \thead[l]{$\theta^k$ \\ not $\theta^k$ \\ }  & \thead[l]{ \\ $a_d^k > a_a^k$ \\ $a_d^k \leq a_a^k$} & \thead[c]{$0.0$ \\ $1.0$ \\ $0.0$} \\
\bottomrule \\[-0.3em]
\multicolumn{4}{c}{Here, we consider $r_a^\theta(s^k, a_d^k, a_a^k)=-r_d^\theta(s^k, a_d^k, a_a^k)$.}
\end{tabular}
\end{center}
\label{tab:IRsetup}
\vspace{-6mm}
\end{table}

\begin{table}[htbp]
\caption{Transition Probability}
\begin{center}
\begin{tabular}{cccccc}
\toprule
$\theta^k$ & $s^k$ & $a_d^k$ & $a_a^k$ & condition & $T^\theta(s^{k+1}| s^k, a_d^k, a_a^k)$ \\
\midrule
$\theta_j$ & $s_l$ & not $\theta_j$ & $\theta_j$ &  & $\alpha-\theta_j/10 - (l/\beta)$ \\
\midrule
$\theta_j$ & $s_l$ & $\theta_j$ & $\theta_j$ &  & $\alpha-\delta-\theta_j/10 - (l/\beta)$ \\
\midrule
$\theta_j$ & $s_l$ & not $\theta_j$ & not $\theta_j$ & \thead[c]{ \\ $a_d^k > a_a^k$ \\ $a_d^k \leq a_a^k$} & \thead[c]{ \\ $\alpha-\delta$ \\ $1.0-(\alpha-\delta)$}  \\
\bottomrule \\[-0.3em]
\multicolumn{6}{c}{$\alpha$ is the attacker's ability, $\delta$ is the defender's ability, and $\beta$ denotes the}\\ 
\multicolumn{6}{c}{impact of the current state. Here, we choose $\alpha=0.8, \delta=0.5, \beta=\infty$.}
\end{tabular}
\end{center}
\label{tab:TPsetup}
\vspace{-6mm}
\end{table}

\begin{table}[htbp]
\caption{Terminal Rewards}
\begin{center}
\begin{tabular}{cccccc}
\toprule
$s^K$ & $s_1$ & $s_2$ & $s_3$ & $s_4$ & $s_5$ \\
\midrule
$U_T(s^K)$ & 100 & 50 & 10 & 0 & -100 \\
\bottomrule \\[-0.3em]
\end{tabular}
\end{center}
\label{tab:TRsetup}
\vspace{-6mm}
\end{table}

From Fig. \ref{fig:v0}, the scenario with no allowed switches and absent deception (case $0$ (mode $0$)) shows that the defender's initial advantage diminishes over time during interactions with the attacker (that is, the attacker finds out that there is no deception), as indicated by the defender's value being the lowest and negative in this case. In the case of no allowed switches with basic deception at the beginning (case $0$ (mode $1$)), the defender can expect to win when the attacker's lifetime is short. However, when the lifetime is long enough ($K=20$), the defender may still lose. The case of no allowed switches with sophisticated deception (case $0$ (mode $2$)) exhibits a similar trend.

Then, we can observe that with a fixed attacker's lifetime, the defender's initial value increases as the budgeted number of switches increases. This indicates that strategically selecting the timing and activation of deception techniques improves the defender's performance. Additionally, we observe that the defender's value at the initial stage decreases as the attacker's lifetime increases in all five cases. This is because a longer attacker's lifetime increases the likelihood of reaching the final state, where critical assets are located. To address this, potential solutions include expanding the set to incorporate more advanced deception techniques, increasing the switch budget, and enhancing intrusion detection techniques to shorten the attacker's lifetime.

\subsection{Comparison between different switching strategies}

\begin{figure}
    \centering
    \includegraphics[width=3.5in]{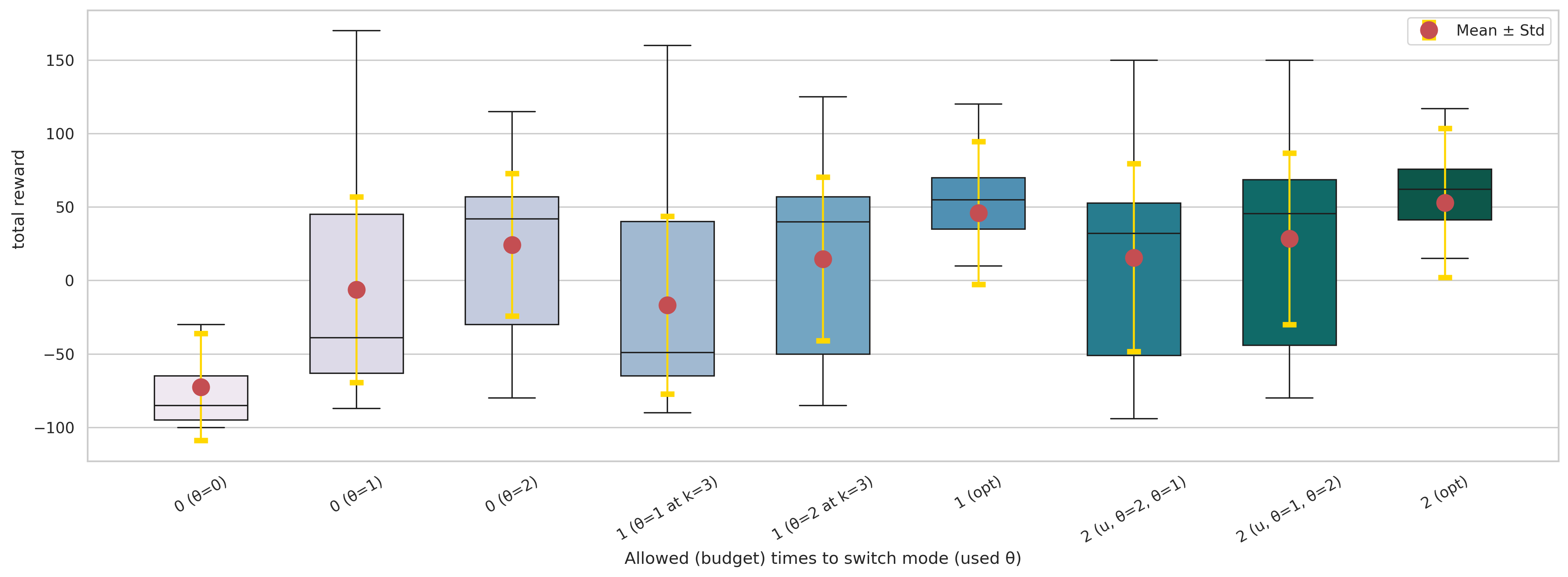}
    \caption{The results for the total rewards (sum of immediate reward at each stage and the terminal reward) under different switching strategies. Here, $K=10$, and each scenario is evaluated over $200$ experiments.}
    \label{fig:arbitrage_}
\vspace{-5mm}
\end{figure}

To illustrate the effectiveness of the optimal switching strategy, we consider the following baselines for comparison purposes. The results for the total rewards (sum of the immediate reward at each stage and the terminal reward) under different scenarios are shown in Fig. \ref{fig:arbitrage_}.

\noindent\textbf{No switch:} In this case, no switches are allowed; the deception technique remains fixed from the initial stage to the terminal stage. For example, in Fig. \ref{fig:arbitrage_}, case ``0 ($\theta=0$)'' denotes a scenario where the budgeted switch times is zero ($M=0$) and the deception is absent (game mode $\theta=0$). 

\noindent\textbf{Fixed timing}: In this case, the timeing to activate the deployed deception (i.e., switch mode) is fixed. For instance, in Fig. \ref{fig:arbitrage_}, case ``1 ($\theta=1$ at $k=3$)'' means that the budgeted switch is one ($M=1$) and that basic deception is activated at stage $3$. 

\noindent\textbf{Uniform interval}: In this case, the time intervals between each deception activation are uniform. For example, in Fig. \ref{fig:arbitrage_}, case ``2 ($\theta=2, \theta=1$)'' indicates that the budgeted switch is two ($M=2$), with sophisticated deception activated first, followed by basic deception. 

\noindent\textbf{Proposed strategy}: In this case, the strategy proposed in Section \ref{sec:optimal_switch} is applied. Cases ``1 (opt)'' and ``2 (opt)'' in Fig. \ref{fig:arbitrage_} indicate that the proposed strategy is used when the budgeted switch times are one ($M=1$) and two ($M=2$), respectively.

From Fig. \ref{fig:arbitrage_}, we observe that when no switch is allowed, deploying sophisticated deception yields the best performance compared to basic or no deception. In the case where one switch is allowed (with the initial game mode $\theta=0$ at stage $k=0$), the proposed strategy achieves the best performance by recommending a switch to sophisticated deception at stage $1$ or $2$. The two fixed-timing cases can be seen as delayed switches, with ``1 ($\theta=1$ at stage $k=3$)'' performing even worse due to selecting the wrong mode compared to the proposed strategy's recommendation. When two switches are allowed, the proposed strategy still outperforms the uniform-interval baseline. This may be because simply applying uniform intervals between switches can lead to unnecessary early activation or delayed responses. As a result, the defender's advantage gained from the attacker's incorrect belief may diminish over time, allowing the attacker to assess the situation (deception, game mode) more correctly and adjust their effort (action) accordingly. Consequently, the attacker avoids being trapped in a specific state and may reach the critical asset.

\section{Conclusion}
In this work, we propose a Stackelberg game-based framework for defensive deception design. Noticing that the defender's initial advantage from deception may diminish during the interaction with the attacker as the attacker adapts, it is essential to dynamically determine both the optimal timing and the selection of deception to sustain the defender’s edge. At the lower tactical layer, a one-sided incomplete information Markov game models the defender-attacker interaction under a given deception. At the upper strategic layer, a stopping-time decision process optimizes the deployment timing and selection of deception to maintain the defender’s superiority and protect critical assets. The numerical results show that while the defender's initial advantage diminishes over time, strategically timed deception techniques can enhance the defender's expected utility and improve total execution rewards.

\bibliography{reference}
\bibliographystyle{IEEEtran}

\end{document}